\documentclass[aps,showpacs,twocolumn]{revtex4}
\usepackage{graphicx}

\begin{document}


\newcommand\be{\begin{equation}}
\newcommand\ba{\begin{eqnarray}}
\newcommand\ee{\end{equation}}
\newcommand\ea{\end{eqnarray}}
\newcommand{\pont}{{\,^\ast\!}R\,R}

\title{Thin accretion disk signatures in dynamical
Chern-Simons modified gravity}

\author{Tiberiu Harko}
\email{harko@hkucc.hku.hk} \affiliation{Department of Physics and
Center for Theoretical and Computational Physics, The University
of Hong Kong, Pok Fu Lam Road, Hong Kong}

\author{Zolt\'{a}n Kov\'{a}cs}
\email{zkovacs@mpifr-bonn.mpg.de} \affiliation{Department of Physics and
Center for Theoretical and Computational Physics, The University
of Hong Kong, Pok Fu Lam Road, Hong Kong}

\author{Francisco S. N. Lobo}
\email{flobo@cii.fc.ul.pt} \affiliation{Centro de F\'{i}sica
Te\'{o}rica e Computacional, Faculdade de Ci\^{e}ncias da
Universidade de Lisboa, Campo Grande, Ed. C8 1749-016 Lisboa,
Portugal}

\date{\today}

\begin{abstract}

A promising extension of general relativity is Chern-Simons (CS)
modified gravity, in which the Einstein-Hilbert action is modified
by adding a parity-violating CS term, which couples to gravity via
a scalar field. In this work, we consider the interesting, yet
relatively unexplored, dynamical formulation of CS modified
gravity, where the CS coupling field is treated as a dynamical
field, endowed with its own stress-energy tensor and evolution
equation. We consider the possibility of observationally testing
dynamical CS modified gravity by using the accretion disk
properties around slowly-rotating black holes. The energy flux,
temperature distribution, the emission spectrum as well as the
energy conversion efficiency are obtained, and compared to the
standard general relativistic Kerr solution. It is shown that the
Kerr black hole provide a more efficient engine for the
transformation of the energy of the accreting mass into radiation
than their slowly-rotating counterparts in CS modified gravity.
Specific signatures appear in the electromagnetic spectrum, thus
leading to the possibility of directly testing CS modified gravity
by using astrophysical observations of the emission spectra from
accretion disks.

\end{abstract}


\maketitle

\section{Introduction}

Recently, modified theories of gravity have received a
considerable amount of attention mainly motivated by the problems
of dark energy (see \cite{darkenergy} for reviews) and dark matter
\cite{darkmatter}, and from quantum gravity. A promising extension
of general relativity is Chern-Simons (CS) modified gravity
\cite{Lue:1998mq,Jackiw:2003pm,Alexander:2009tp}, in which the
Einstein-Hilbert action is modified by adding a parity-violating
CS term, which couples to gravity via a scalar field. It is
interesting to note that the CS correction introduces a means to
enhance parity violation through a pure curvature term, as opposed
to through the matter term, as is usually considered in general
relativity. In fact, CS modified gravity can be obtained
explicitly from superstring theory, where the CS term in the
Lagrangian density is essential due to the Green-Schwarz
anomaly-canceling mechanism, upon four-dimensional
compactification \cite{4dim-comp}.
Two formulations of CS modified gravity exist as independent
theories, namely, the nondynamical formulation and the dynamical
formulation (see \cite{Alexander:2009tp} for an excellent recent
review). In the former, the CS scalar is an {\it a priori}
prescribed function, where its effective evolution equation
reduces to a differential constraint on the space of allowed
solutions; in the latter, the CS is treated as a dynamical field,
possessing an effective stress-energy tensor and an evolution
equation. The majority of the work, up to date, has considered the
nondynamical formulation
\cite{nondyn-form,rotBHnondyn1a,rotBHnondyn1b,rotBHnondyn2},
whereas the dynamical formulation remains mostly unexplored
territory.

Relative to rotating black hole spacetimes, several solutions in
the nondynamical formulation were found in CS modified gravity
\cite{rotBHnondyn1a,rotBHnondyn1b,rotBHnondyn2}. The first
solutions were found by Alexander and Yunes
\cite{rotBHnondyn1a,rotBHnondyn1b}, using a far-field
approximation (where the field point distance is considered to be
much larger than the black hole mass). The second rotating black
hole solution was found by Konno {\it et al} \cite{rotBHnondyn2},
using a small slow-rotation approximation, where the spin angular
momentum is assumed to be much smaller than the black hole mass.
However, it is interesting to note that recently, using the
dynamical formulation of CS modified gravity, spinning black hole
solutions in the slow-rotation approximation have been obtained
\cite{CSslowrot,CSslowrot2}.

An interesting feature of CS modified gravity is that it has a
characteristic observational signature, which could allow to
discriminate an effect of this theory from other phenomena.
However, most of the tests of CS modified gravity to date have
been performed with astrophysical observations and concern the
non-dynamical framework. In particular, it was found that the CS
modified theory predicts an anomalous precession effect
\cite{tests}, which was tested \cite{tests2} with LAGEOS
\cite{lageos}. Another constraint on the non-dynamical theory was
proposed in \cite{rot-curves}, where it was considered that the CS
correction could be used to explain the flat rotation curves of
galaxies. However, in \cite{Yunes-Spergel} a bound was placed on
the non-dynamical model with a canonical CS scalar that is eleven
orders of magnitude stronger than the Solar System one, using
double binary pulsar data. Recently, using the dynamical
formulation of CS modified gravity, a stringent constraint was
placed on the coupling parameter associated to the dynamical
coupling of the scalar field \cite{Cardoso:2009pk}.

In this work, we further extend the constraints placed on the
dynamical formulation of CS gravity by using the observational
signatures of thin disk properties around rotating black holes. In
the context of stationary axisymmetric spacetimes, the mass
accretion around rotating black holes was studied in general
relativity for the first time in \cite{NoTh73}, by extending the
theory of non-relativistic accretion \cite{ShSu73}.  The radiation
emitted by the disk surface was also studied under the assumption
that black body radiation would emerge from the disk in
thermodynamical equilibrium \cite{PaTh74,Th74}. More recently, the
emissivity properties of the accretion disks were investigated for
exotic central objects, such as wormholes \cite{Harko:2008vy}, and
non-rotating or rotating quark, boson or fermion stars,
brane-world black holes or gravastars
\cite{Bom,To02,YuNaRe04,Guzman:2005bs,Pun:2008ua,KoChHa09,grav,
KoChHa091,Pun:2008ae, Harko:2009rp}.

Thus, it is the purpose of the present paper to study the thin
accretion disk models for slowly-rotating black holes in the
dynamical formulation of CS modified theories of gravity, and
carry out an analysis of the properties of the radiation emerging
from the surface of the disk. As compared to the standard general
relativistic case, significant differences appear in the energy
flux and electromagnetic spectrum for CS slowly-rotating black
holes, thus leading to the possibility of directly testing CS
modified gravity by using astrophysical observations of the
emission spectra from accretion disks.

The present paper is organized as follows. In Sec.~\ref{sec:II},
we review the dynamical formulation of CS modified gravity, and
present the Yunes-Pretorius (YP) slowly-rotating solution found in
\cite{CSslowrot2}. In Sec.~\ref{sec:III}, we review the formalism
and the physical properties of the thin disk accretion onto
compact objects, for stationary axisymmetric spacetimes. In
Sec.~\ref{sec:IV}, we analyze the basic properties of matter
forming a thin accretion disk in slowly-rotating black hole
spacetimes in CS modified gravity. We discuss and conclude our
results in Sec.~\ref{sec:concl}. Throughout this work, we use a
system of units so that $c=G=\hbar =k_{B}=1$, where $k_{B}$ is
Boltzmann's constant.

\section{Dynamical Chern-Simons modified gravity}\label{sec:II}

In this Section, we write down the field equations of the
Chern-Simons gravity, and present the Yunes-Pretorius (YP)
slowly-rotating solution found in \cite{CSslowrot2}.

\subsection{Field equations of Chern-Simons theory}

Consider the dynamical Chern-Simons modified gravity theory
provided by the action in the form
\begin{equation}
S = S_{\rm EH} + S_{\rm CS} +  S_{\vartheta} + S_{\rm mat} \,.
\label{CSaction}
\end{equation}
The first term is the standard Einstein-Hilbert action
\begin{equation} \label{EH-action}
S_{\rm{EH}} = \kappa \int d^4x \sqrt{-g} R\,,
\end{equation}
where $\kappa^{-1} = 16 \pi G$ and $R$ is the Ricci scalar. The
second term defined as
\begin{equation} \label{CS-action}
S_{\rm{CS}} = \frac{\alpha}{4} \int d^4x
\sqrt{-g} \; \vartheta \; \pont\,,
\end{equation}
is the Chern-Simons correction; the third term
\begin{equation} \label{Theta-action}
S_{\vartheta} = - \frac{\beta }{2} \int d^{4}x \sqrt{-g} \left[
g^{\mu\nu} \left(\nabla_{\mu} \vartheta\right) \left(\nabla_{\nu}
\vartheta\right) + 2 V(\vartheta) \right]\,,
\end{equation}
is the scalar field term. The matter action is given by
\begin{equation}
S_{\rm{mat}} = \int d^{4}x \sqrt{-g} {\cal{L}}_{\rm{mat}}\,,
\end{equation}
where ${\cal{L}}_{\rm{mat}}$ the matter Lagrangian.

The parameters $\alpha$ and $\beta$ are dimensional coupling
constants; the {\it{CS coupling field}}, $\vartheta$, is a
function of spacetime that parameterizes deformations from GR
\cite{CSslowrot2}; $\nabla_{\mu}$ is the covariant derivative
associated with the metric tensor $g_{\mu\nu}$; and the quantity
$^\ast R R$ is the Pontryagin density defined as
\begin{equation}
^\ast RR = \, ^\ast
R^{\tau}{}_{\sigma}{}^{\mu\nu}R^{\sigma}{}_{\tau\mu\nu} \,,
\label{Pontryagin}
\end{equation}
where the dual Riemann tensor is given by $^\ast
R^{\tau}{}_{\sigma}{}^{\mu\nu}=\frac{1}{2}\epsilon^{\mu\nu\alpha\beta}
R^{\tau}{}_{\sigma\alpha\beta}$, with
$\epsilon^{\mu\nu\alpha\beta}$ the 4-dimensional Levi-Civita
tensor.

Varying the action $S$ with respect to the metric $g_{\mu\nu}$ one
obtains the gravitational field equation given by
\begin{eqnarray}
G_{\mu\nu}+\frac{\alpha}{\kappa}\,C_{\mu\nu}=\frac{1}{2\kappa}
\left( T^{\rm mat}_{\mu\nu}+T^\vartheta_{\mu\nu}\right)\,,
    \label{modfieldeq}
\end{eqnarray}
where $G_{\mu\nu}$ is the Einstein tensor, and $C_{\mu\nu}$ is the
cotton tensor defined as
\begin{eqnarray}
C^{\mu\nu}=\nabla_\sigma\vartheta\;\epsilon^{\sigma\alpha\beta(\mu}
\nabla_\beta\,R^{\nu)}{}_{\alpha}
+\nabla_\sigma\nabla_\alpha\vartheta\;^\ast
R^{\alpha(\mu\nu)\sigma}\,.
    \label{Cotton}
\end{eqnarray}

The total stress-energy tensor is split into the matter term
$T_{\rm mat}^{\mu\nu}$, and the scalar field contribution
$T_\vartheta^{\mu\nu}$, which is provided by the following
relationship
\begin{equation}
T_{\mu\nu}^{\vartheta} =   \beta  \left[ \left(\nabla_{\mu}
\vartheta\right) \left(\nabla_{\nu} \vartheta\right)- \frac{1}{2}
g_{\mu\nu}\left(\nabla_{\mu} \vartheta\right) \left(\nabla^{\nu}
\vartheta\right)-g_{\mu\nu}  V(\vartheta)  \right]\,.
    \label{stress-scalar}
\end{equation}

Varying the action with respect to the scalar field $\vartheta$,
one obtains the equation of motion for the Chern-Simons coupling
term, given by
\begin{equation}
\beta \; \nabla_\mu \nabla^\mu\vartheta = \beta \;
\frac{dV}{d\vartheta} - \frac{\alpha}{4} \pont\,.
    \label{EOMscalar}
\end{equation}
Note that the evolution of the CS coupling is not only governed by
its stress-energy tensor, but also by the curvature of spacetime.
In the nondynamical formulation of CS modified gravity the
constraint $\beta=0$ is considered, while in the dynamical
framework, $\beta$ is allowed to be arbitrary, so that Eq.
(\ref{EOMscalar}) is now the evolution equation for the CS
coupling field.

Considering the diffeomorphism invariance of the matter part of
the action, we have $\nabla_{\mu} T^{\mu\nu}_{\rm{mat}} = 0$, and
taking into account the Bianchi identities, i.e., $\nabla_{\mu}
G^{\mu\nu}=0$, provides the following conservation law
\begin{equation} \label{nablaC}
\nabla_\mu C^{\mu\nu} = - \frac{1}{8} (\nabla^\nu \vartheta)
\pont.
\end{equation}


\subsection{Rotating black hole solutions in Chern-Simons model}

In this paper, we consider the Yunes-Pretorius (YP)
slowly-rotating solution found in \cite{CSslowrot2}, where the CS
correction provides an effective reduction of the frame-dragging
around a black hole, in comparison with that of the Kerr solution.
We will not analyze the Konno, Matsuyama and Tanda (KMT)
approximation \cite{CSslowrot}, since the YP solution is taken to second order in the rotation parameter, and therefore gives more accurate results.

The Yunes-Pretorius (YP) approximation method employs two schemes
\cite{CSslowrot2}, namely, a small-coupling approximation and a
slow-rotation approximation. In particular, the small-coupling
scheme treats the CS modification as a small deformation of
general relativity. The slow-rotation scheme expands the
background perturbations in powers of the Kerr rotation parameter
$a$, and the background metric is formalized via the Hartle-Thorne
approximation \cite{Hartle-Thorne}. We refer the reader to
\cite{CSslowrot2} for details, and present the final metric given
by
\begin{eqnarray}
ds^2&=&-\left[\left(1-\frac{2M}{r}\right)+\frac{2a^2M}{r^3}
\cos^2\theta\right]\,dt^2\nonumber\\
&& \hspace{-1.25cm}+ \left(1-\frac{2M}{r}\right)^{-1}
\left[1+\frac{a^2}{r^2}\left(\cos^2\theta-
\left(1-\frac{2M}{r}\right)^{-1}\right)\right]\:dr^2
    \nonumber\\
&& \hspace{-1.25cm}- \left[\frac{4 M a}{r} \sin^{2}{\theta}\right.
- \left.\frac{10}{8} \frac{\xi a}{r^{4}} \left(1 + \frac{12 M}{7 r}
+ \frac{27 M^{2}}{10 r^{2}} \right) \sin^{2}{\theta}\right] dtd\phi
\nonumber \\
& & + \left(r^{2} + a^{2} \cos^{2}{\theta}\right)d\theta^2
   \nonumber \\
& & + \left[r^{2} \sin^{2}{\theta} + a^{2} \sin^{2}{\theta}
\left(1 + \frac{2 M}{r} \sin^{2}{\theta}
\right)\right]d\phi^2\:,\label{rot-metric2}
\end{eqnarray}
with $\xi=\alpha^2/(\kappa\beta)$.

In the following, we compare the properties of the metric given by
Eq.~(\ref{rot-metric2}) with the standard general relativistic
Kerr metric, respectively, which in the equatorial approximation
can be written as \cite{PaTh74}
\begin{equation}
ds^2=-DA^{-1}dt^2+r^2A\left(d\phi -\omega dt\right)^2+D^{-1}dr^2+dz^2,
\end{equation}
where the coordinate $z=r\cos\theta $ was used to replace $\theta
$, and $A=1+a_{*}^2x^{-4}+2a_{*}^2x^{-6}$ and
$D=1-2x^{-2}+a_{*}^2x^{-4}$, respectively. The dimensionless
coordinate $x$ is defined as $x=\sqrt{r/M}$, and the spin
parameter $a_{*}$ is defined as $a_{*}=J/M^{2}=a/M$.

\section{Thermal equilibrium radiation properties of thin
accretion disks in stationary axisymmetric
spacetimes}\label{sec:III}

\subsection{Stationary and axially symmetric spacetimes}

The physical properties and the electromagnetic radiation
characteristics of particles moving in circular orbits around
general relativistic bodies are determined by the geometry of the
spacetime around the compact object.  For a stationary and axially
symmetric geometry the metric is given in a general form by
\begin{equation}\label{rotmetr1}
ds^2=g_{tt}\,dt^2+2g_{t\phi}\,dt d\phi+g_{rr}\,dr^2
+g_{\theta\theta}\,d\theta^2+g_{\phi\phi}\,d\phi^2\,.
\end{equation}
In the equatorial approximation, which is the case of interest for
our analysis, the metric functions $g_{tt}$, $g_{t\phi}$,
$g_{rr}$, $g_{\theta\theta}$ and $g_{\phi\phi}$ only depend on the
radial coordinate $r$, i.e., $|\theta-\pi /2|\ll 1$.

To compute the relevant physical quantities of thin accretion
disks, we determine first the radial dependence of the angular
velocity $\Omega $, of the specific energy $\widetilde{E}$, and of
the specific angular momentum $\widetilde{L}$ of particles moving
in circular orbits in a stationary and axially symmetric geometry
through the geodesic equations. The latter take the following form
\cite{Harko:2008vy}
\begin{eqnarray}
\frac{dt}{d\tau}&=&\frac{\widetilde{E}
g_{\phi\phi}+\widetilde{L}g_{t\phi}}{g_{t\phi}^2-g_{tt}g_{\phi\phi}},
   \label{geodeqs1}   \\
\frac{d\phi}{d\tau}&=&-\frac{\widetilde{E}
g_{t\phi}+\widetilde{L}g_{tt}}{g_{t\phi}^2-g_{tt}g_{\phi\phi}},
    \label{geodeqs2}  \\
g_{rr}\left(\frac{dr}{d\tau}\right)^2&=&-1+\frac{\widetilde{E}^2
g_{\phi\phi}+2\widetilde{E}\widetilde{L}g_{t\phi}
+\widetilde{L}^2g_{tt}}{g_{t\phi}^2-g_{tt}g_{\phi\phi}}.
    \label{geodeqs3}
\end{eqnarray}
From Eq.~(\ref{geodeqs3}) one can introduce an effective potential
term as
\begin{equation}\label{roteffpot}
V_{eff}(r)=-1+\frac{\widetilde{E}^2
g_{\phi\phi}+2\widetilde{E}\widetilde{L}g_{t\phi}
+\widetilde{L}^2g_{tt}}{g_{t\phi}^2-g_{tt}g_{\phi\phi}}.
\end{equation}

For stable circular orbits in the equatorial plane the following
conditions must hold: $V_{eff}(r)=0$ and $V_{eff,\;r}(r)=0$, where
the comma in the subscript denotes a derivative with respect to
the radial coordinate $r$. These conditions provide the specific
energy, the specific angular momentum and the angular velocity of
particles moving in circular orbits for the case of spinning
general relativistic compact spheres, given by
\begin{eqnarray}
\widetilde{E}&=&-\frac{g_{tt}+g_{t\phi}\Omega}{\sqrt{-g_{tt}
-2g_{t\phi}\Omega-g_{\phi\phi}\Omega^2}},
    \label{rotE}  \\
\widetilde{L}&=&\frac{g_{t\phi}+g_{\phi\phi}\Omega}{\sqrt{-g_{tt}
-2g_{t\phi}\Omega-g_{\phi\phi}\Omega^2}},
     \label{rotL}  \\
\Omega&=&\frac{d\phi}{dt}=\frac{-g_{t\phi,r}+\sqrt{(g_{t\phi,r})^2
-g_{tt,r}g_{\phi\phi,r}}}{g_{\phi\phi,r}}.
     \label{rotOmega}
\end{eqnarray}
The marginally stable orbit around the central object can be
determined from the further condition  $V_{eff,\;rr}(r)=0$,
which provides the following
important relationship
\begin{eqnarray}
0& = & (g_{t\phi}^2-g_{tt}g_{\phi\phi})V_{eff,rr}
   \nonumber\\
 & = &  \widetilde{E}^{2}g_{\phi\phi,rr}
 +2\widetilde{E}\widetilde{L}g_{t\phi,rr} +\widetilde{L}^{2}g_{tt,rr}
 \nonumber \\
& & -(g_{t\phi}^{2} -g_{tt}g_{\phi\phi})_{,rr}\;,\label{mso-r}
\end{eqnarray}
where $g_{t\phi}^2-g_{tt}g_{\phi\phi}$ (appearing as a cofactor in
the metric determinant) never vanishes. By inserting
Eqs.~(\ref{rotE})-(\ref{rotOmega}) into Eq.~(\ref{mso-r}) and
solving this equation for $r$, we obtain the radii of the
marginally stable orbits, once the metric coefficients $g_{tt}$,
$g_{t\phi}$ and $g_{\phi\phi}$ are explicitly given.

\subsection{Physical properties of thin accretion disks}

For the thin accretion disk, it is assumed that its vertical
size is negligible, as compared to its horizontal extension, i.e,
the disk height $H$, defined by the maximum half thickness of the
disk, is always much smaller than the characteristic radius $r$ of
the disk, $H \ll r$.  The thin disk has an inner edge at the
marginally stable orbit of the compact object potential, and the
accreting plasma has a Keplerian motion in higher orbits.

In steady state accretion disk models, the mass accretion rate
$\dot{M}_{0}$ is assumed to be a constant that does not change
with time.  The radiation flux $F$ emitted by the surface of the
accretion disk can be derived from the conservation equations for
the mass, energy and angular momentum, respectively. Then the
radiant energy $F(r)$ over the disk is expressed in terms of the
specific energy, of the angular momentum, and of the angular
velocity of the particles orbiting in the disk
\cite{NoTh73,PaTh74},
\begin{equation}
F(r)=-\frac{\dot{M}_{0}}{4\pi \sqrt{-g}}\frac{\Omega
_{,r}}{(\widetilde{E}-\Omega
\widetilde{L})^{2}}\int_{r_{ms}}^{r}(\widetilde{E}-\Omega
\widetilde{L}) \widetilde{L}_{,r}dr\;,  \label{F}
\end{equation}
where $\dot{M}_0$ is the mass accretion rate, measuring the rate at
which the rest mass of the particles flows inward through the disk
with respect to the coordinate time $t$ and $r_{ms}$ is the
marginally stable orbit obtained from Eq.~(\ref{mso-r}).

Another important characteristics of the mass accretion process is
the efficiency with which the central object converts rest mass
into outgoing radiation. This quantity is defined as the ratio of
the rate of the radiation energy of photons, escaping from the
disk surface to infinity, and the rate at which mass-energy is
transported to the central compact general relativistic object,
both measured at infinity \cite{NoTh73,PaTh74}. If all the emitted
photons can escape to infinity, the efficiency is given in terms
of the specific energy measured at the marginally stable orbit $
r_{ms}$,
\begin{equation}
\epsilon =1-\widetilde{E}_{ms}.  \label{epsilon}
\end{equation}
For Schwarzschild black holes the efficiency $\epsilon $ is about
$6\%$, whether the photon capture by the black hole is considered,
or not. Ignoring the capture of radiation by the black hole, $\epsilon $
is found to be $42\%$ for rapidly rotating black holes, whereas
the efficiency is $40\%$ with photon capture in the Kerr potential
\cite{Th74}.

The accreting matter in the steady-state thin disk model is
supposed  to be in thermodynamical equilibrium. Therefore the
radiation emitted by the disk surface can be considered as a
perfect black body radiation, where the energy flux is given by
$F(r)=\sigma T^{4}(r)$ ($\sigma $ is the Stefan-Boltzmann
constant), and the observed luminosity $L\left( \nu \right)$ has a
redshifted black body spectrum \citep{To02}:
\begin{equation}
L\left( \nu \right) =4\pi d^{2}I\left( \nu \right)
=\frac{8}{\pi }\cos
\gamma \int_{r_{i}}^{r_{f}}\int_0^{2\pi}\frac{\nu^{3}_e r
d\phi dr }{\exp
\left(\nu_e/T\right) -1}.\label{L}
\end{equation}

Here $d$ is the distance to the source, $I(\nu )$ is the thermal
energy flux radiated by the disk, $\gamma $ is the disk
inclination angle, and $r_{i}$ and $r_{f}$ indicate the position
of the inner and outer edge of the disk, respectively. We take
$r_{i}=r_{ms}$ and $r_{f}\rightarrow \infty $, since we expect the
flux over the disk surface vanishes at $r\rightarrow \infty $ for
any kind of general relativistic compact object geometry. The
emitted frequency is given by $\nu_e=\nu(1+z)$, and the redshift
factor can be written as
\begin{equation}
1+z=\frac{1+\Omega r \sin \phi \sin \gamma }{\sqrt{ -g_{tt}
- 2 \Omega g_{t\phi} - \Omega^2 g_{\phi\phi}}},
\end{equation}
where we have neglected the light bending \citep{Lu79,BMT01}.

\section{Electromagnetic signatures of accretion
disks around slowly-rotating black holes in dynamical Chern-Simons
gravity}\label{sec:IV}

Close to the equatorial plane of the slowly-rotating black holes,
one can introduce the coordinate $z=r\cos\theta$ describing ``the
height above the equatorial plane'' and write the metrics given by
Eq.~(\ref{rot-metric2}) in the form
 \begin{eqnarray} ds^{2}&=&-f(r)dt^{2}-4\frac{Ma}{r}[1+h_{1}(r)]dtd\phi
 +\frac{1+h_{2}(r)}{f(r)}dr^{2}
   \nonumber\\
& & +r^{2}[1+h_{3}(r)]d\phi^{2}+dz^{2}\:,\label{rot-met-gen}
\end{eqnarray}
where $f(r)=1-2M/r$ is the Schwarzschild form factor, and
\begin{eqnarray}
h_{1}(r) & = & -\frac{5}{16}\frac{\xi}{Mr^{3}}\left(1+\frac{12}{7}
\frac{M}{r}+\frac{27}{10}\frac{M^{2}}{r^{2}}\right)\:,\label{eq:h1YP}
    \\
h_{2}(r) & = & -\frac{a^{2}}{r^{2}f(r)}\;,\label{eq:h2YP}
      \\
h_{3}(r) & = & \frac{a^{2}}{r^{2}}\left(1+\frac{2M}{r}\right)
\label{eq:h3YP}
\end{eqnarray}
for the YP metric.

\subsection{Constants of motion}

If we insert the metric components of Eq.~(\ref{rot-met-gen}) into
the expressions (\ref{rotE})-(\ref{rotOmega}) of the specific
energy, of the specific angular momentum, and of the angular
velocity, we obtain
\begin{eqnarray}
\widetilde{E}&=&\frac{f+2Mar^{-1}\Omega(1+h_1)}{\sqrt{f+4Mar^{-1}
\Omega(1+h_1) - r^2\Omega^2(1+h_3)}}\:,\label{rotE2}\\
\widetilde{L}&=&\frac{-2Mar^{-1}(1+h_1)+r^2\Omega(1+h_3)}
{\sqrt{f+4Mar^{-1}\Omega(1+h_1) - r^2\Omega^2(1+h_3)}}\:,\\
\Omega&=& -\frac{Ma}{2r^3}\left[H_1(r)\pm\sqrt{H_1(r)+\frac{r^3}
{Ma^2H_2(r)}}
\right],\label{rotOmega2}
\end{eqnarray}
with
\begin{eqnarray*}
H_1(r)&=&\frac{1+h_1-rh_{1,r}}{H_2(r)},\\
H_2(r)&=&\frac{1}{2}(1+h_3+rh_{3,r})\:.\\
\end{eqnarray*}

As Eqs.~(\ref{rotE2})-(\ref{rotOmega2}) show, the constants of
motion for the particles orbiting in the equatorial plane depend
only on the metric functions $f(r)$, $h_{1}(r)$ and $h_{3}(r)$,
respectively. The coupling constant $\xi$ of the CS gravity
appears only in $h_{1}(r)$. Since $h_{3}(r)$ is non zero, it
reduces the value of $g_{\phi\phi}$ for any $r$ outside the
marginally stable orbit. However, the decrease in $g_{\phi\phi,r}$
is proportional to $h_{3,r}\sim a^{2}r^{-3}$, and it causes only a
small variation in the radial distribution of the angular
velocity. The specific energy and angular momentum, depending on
$g_{\phi\phi}$, decreases only slightly in amplitude as well. The
non-vanishing functions $h_{1}(r),$ $h_{2}(r)$ and $h_{3}(r)$ give
a negligible contribution to the volume element
 \[
\sqrt{-g_{CS}}=\sqrt{(1+h_{2})\left[\frac{4M^{2}a^{2}
(1+h_{1})^{2}}{r^{2}-2Mr}+r^{2}(1+h_{3})\right]},
\]
as compared to the case of the equatorial approximation for slowly
rotating general relativistic black holes,
\[\sqrt{-g_{GR}}=\sqrt{\frac{4M^{2}a^{2}}{r^{2}-2Mr}+r^{2}}.\]

As a result, the properties of particles orbiting the equatorial
plane of slowly-rotating black holes in the standard general
relativistic theory and in CS modified gravity are essentially the
same. The only difference is in the location of the marginally
stable orbits, which are strongly affected by the coupling. Since
the inner edge of a thin accretion disk is supposed to be at the
radius $r_{ms}$, the radial profile of the energy flux radiated
over the disk surface can indicate the differences in the mass
accretion processes in the general relativistic theory, and in its
CS type modification, respectively.

\subsection{Flux and temperature distribution}

\begin{figure*}
\centering
\includegraphics[width=2.75in]{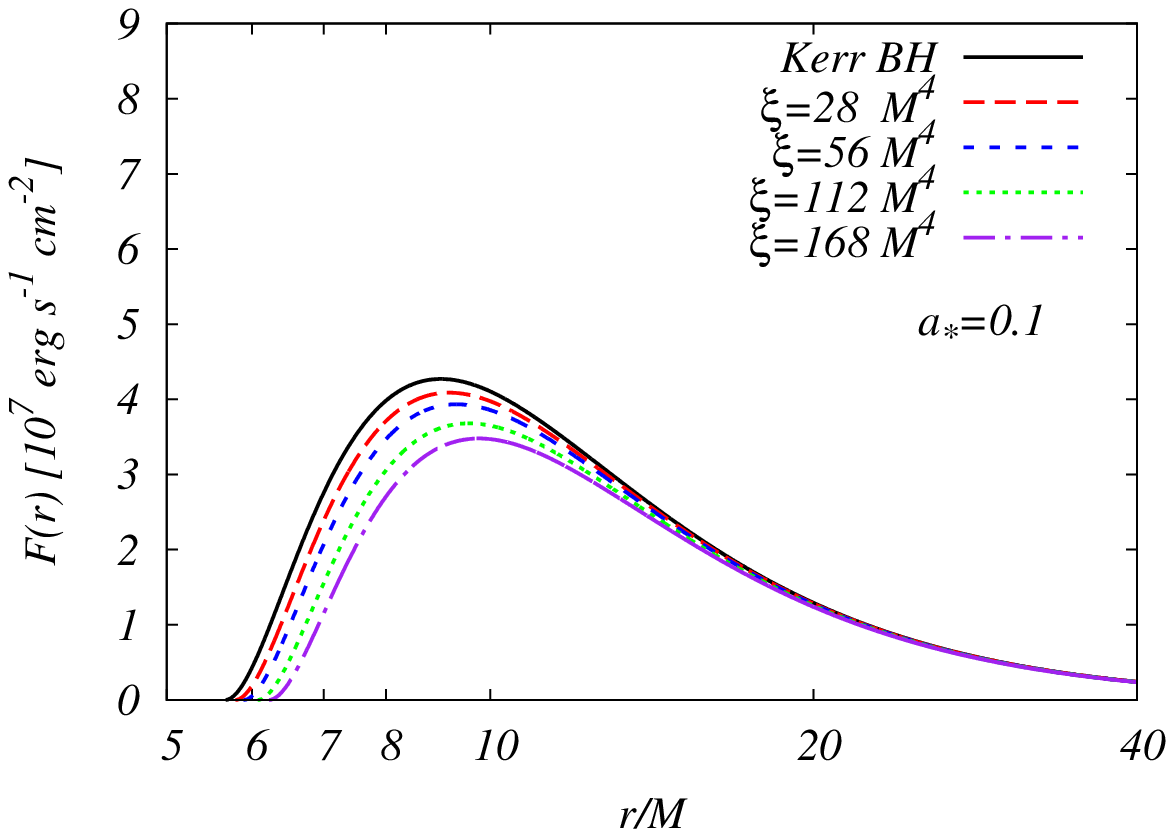}
\hspace{0.2in}
\includegraphics[width=2.75in]{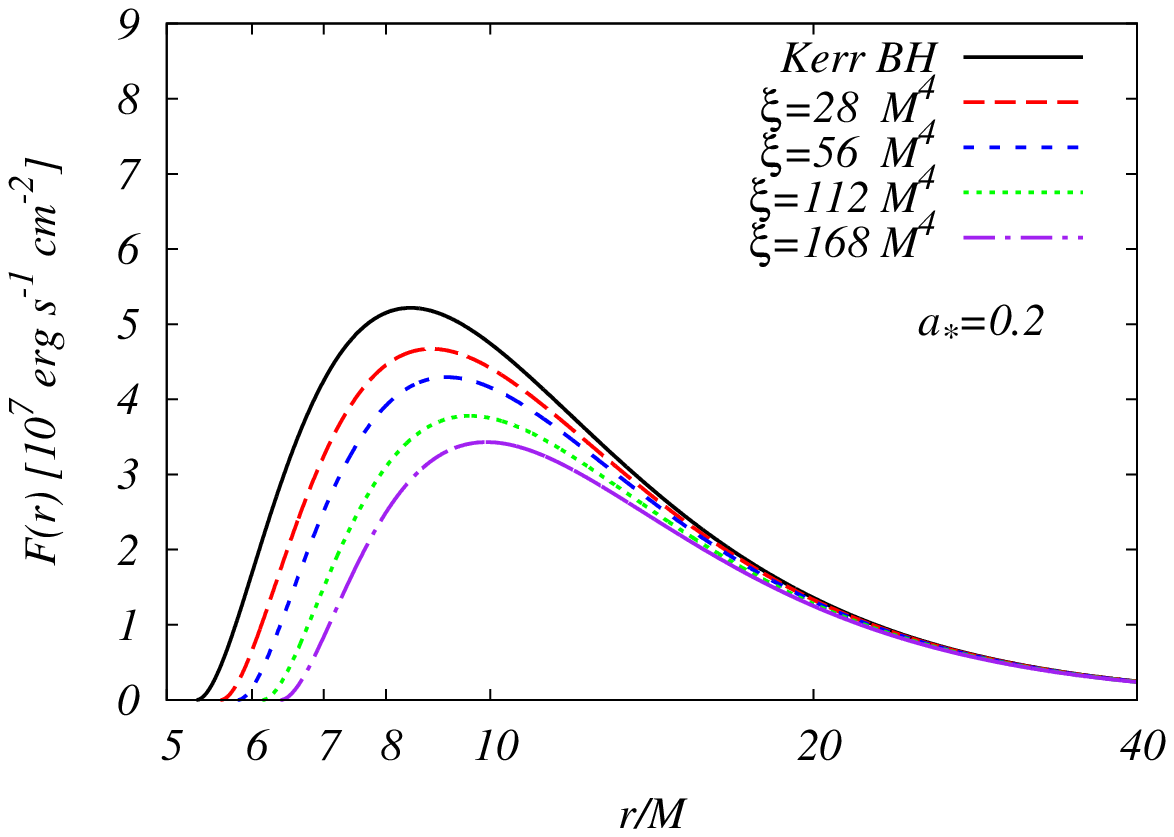}
\hspace{0.2in}
\includegraphics[width=2.75in]{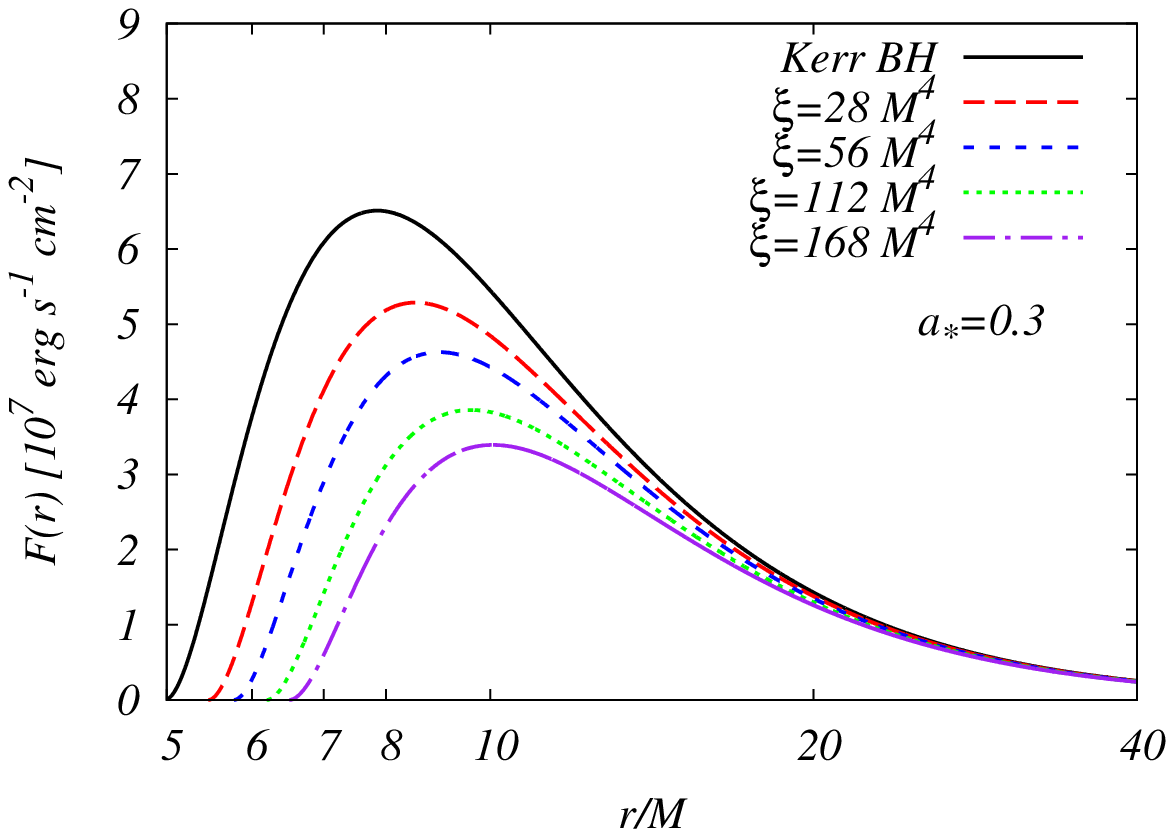}
\hspace{0.2in}
\includegraphics[width=2.75in]{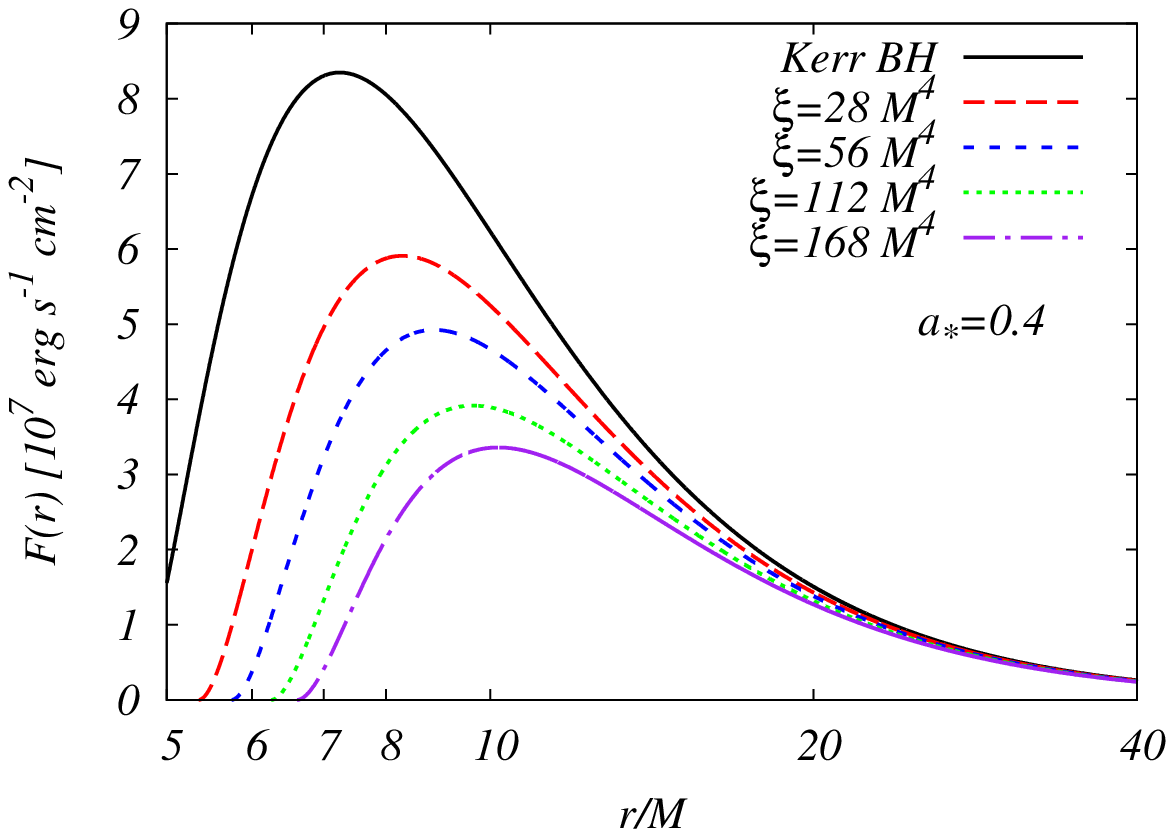}
\caption{The energy flux from accretion disks around
slowly-rotating black holes for different spin parameters in the
general relativity, and in the modified CS theory of gravity with
the YP approximation. The coupling constant $\xi$ is running from
$28M^{4}$ to $168M^{4}$. The spin parameters are set to
$a_*=0.1$ (upper left hand plot), $a_*=0.2$ (upper right hand plot),
$a_*=0.3$ (lower left hand plot) and $a_*=0.4$ (lower right hand plot),
respectively. The total black hole mass is $10^6 M_{\odot}$ and
the mass accretion rate is $10^{-12} M_{\odot}$/year=$4.22
\times10^{-10}\dot{M}_{Edd}$.}
 \label{Fig:flux2}
\end{figure*}

In Figs.~\ref{Fig:flux2} we present the flux distribution for the
slowly-rotating Kerr black holes, and for the slowly rotating YP
solution, respectively. We consider the mass accretion driven by
black holes with a total mass of $M=10^{6}M_{\odot}$, and with a
mass accretion rate of $\dot{M}_{0}=10^{-12}M_{\odot}$/year. In
units of the Eddington accretion rate $\dot{M}_{Edd}=1.5\times
10^{17}\left(M/M_{\odot}\right)$ g/s we have
$\dot{M}_{0}=4.22\times 10^{-10}\dot{M}_{Edd}$. The spin parameter
$a_{*}$, runs from 0.1 to 0.4, whereas the coupling constant of
the CS gravity is set to $\xi=28M^{4}$, $56M^{4}$, $112M^{4}$ and
$168M^{4}$, respectively.

The plots show that the energy flux profiles of the disks in the
CS modified gravity models deviate from the slowly-rotating
general relativistic Kerr black hole case. For the smallest values
of $\xi$, the inner edge of the accretion disk is located at
somewhat higher radius than the inner edge of the disk around the
Kerr black hole (the location of the marginally stable orbits can
be found in Table~\ref{Tab:eff2}, respectively, considered below).
As the quantities $\widetilde{E}$, $\widetilde{L}$ and $\Omega $
in the flux integral (\ref{F}) are still close for the CS model of
gravity to those for the general relativistic case, for the lower
boundary $r_{ms,GR}<r_{ms,CS}$, the integral gives lower flux
values. Thus, the maximum of the integrated flux is smaller in the
CS modified theory of gravity, and it decreases further as we
increase the coupling constant. The effect of the coupling becomes
more important as the black holes are rotating faster: for higher
values of the spin parameter the same increase in the coupling
constants produces considerably lower flux values, and shifts the
marginally stable orbit to somewhat higher radii, as compared to
the case of the very slow rotation (like, for example, in  the
cases with $a_{*}=0.1$ and $a_{*}=0.4$).

\begin{figure*}
\centering
\includegraphics[width=2.8in]{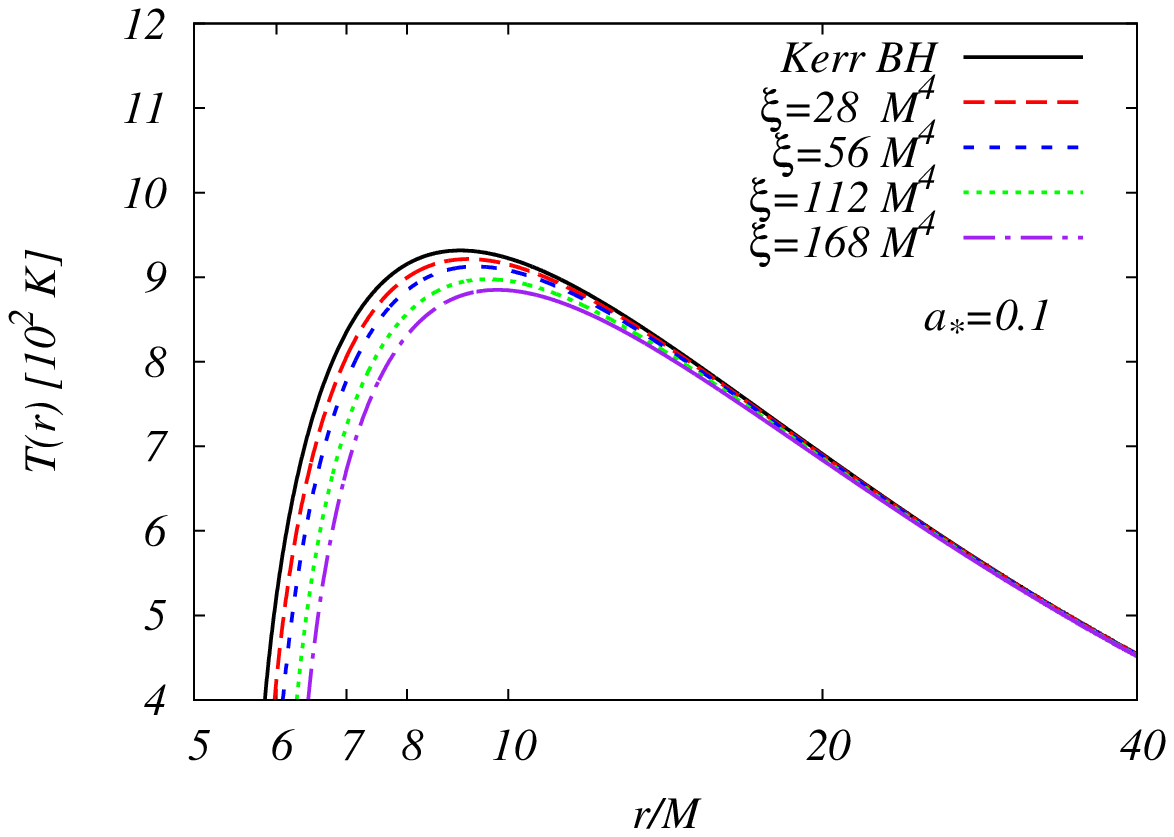}
\hspace{0.2in}
\includegraphics[width=2.8in]{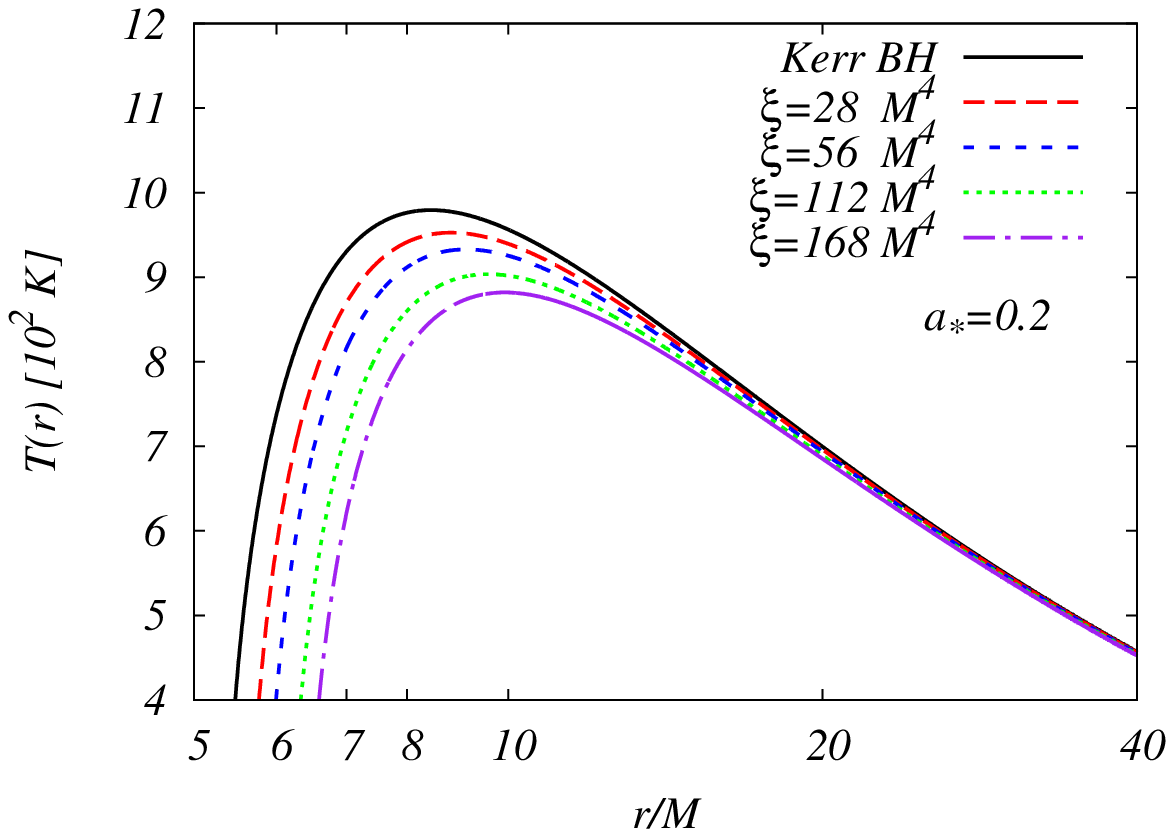}
\hspace{0.2in}
\includegraphics[width=2.8in]{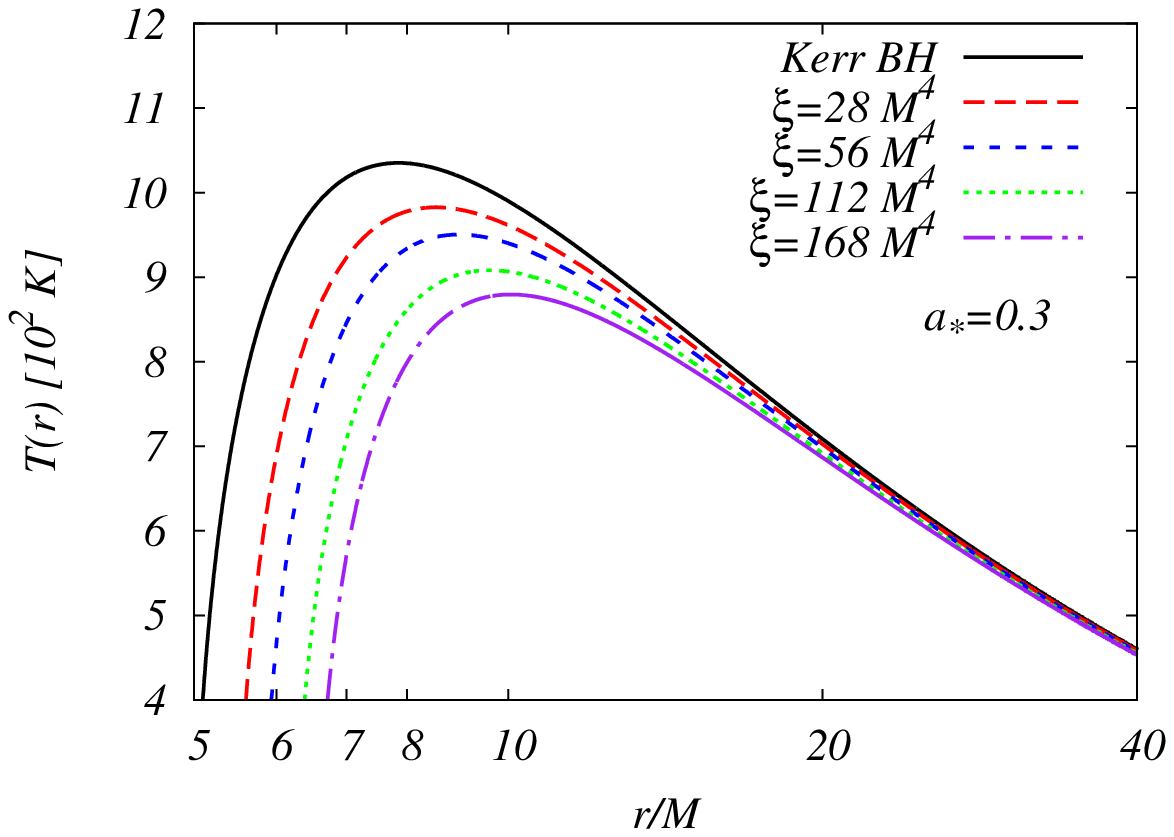}
\hspace{0.2in}
\includegraphics[width=2.8in]{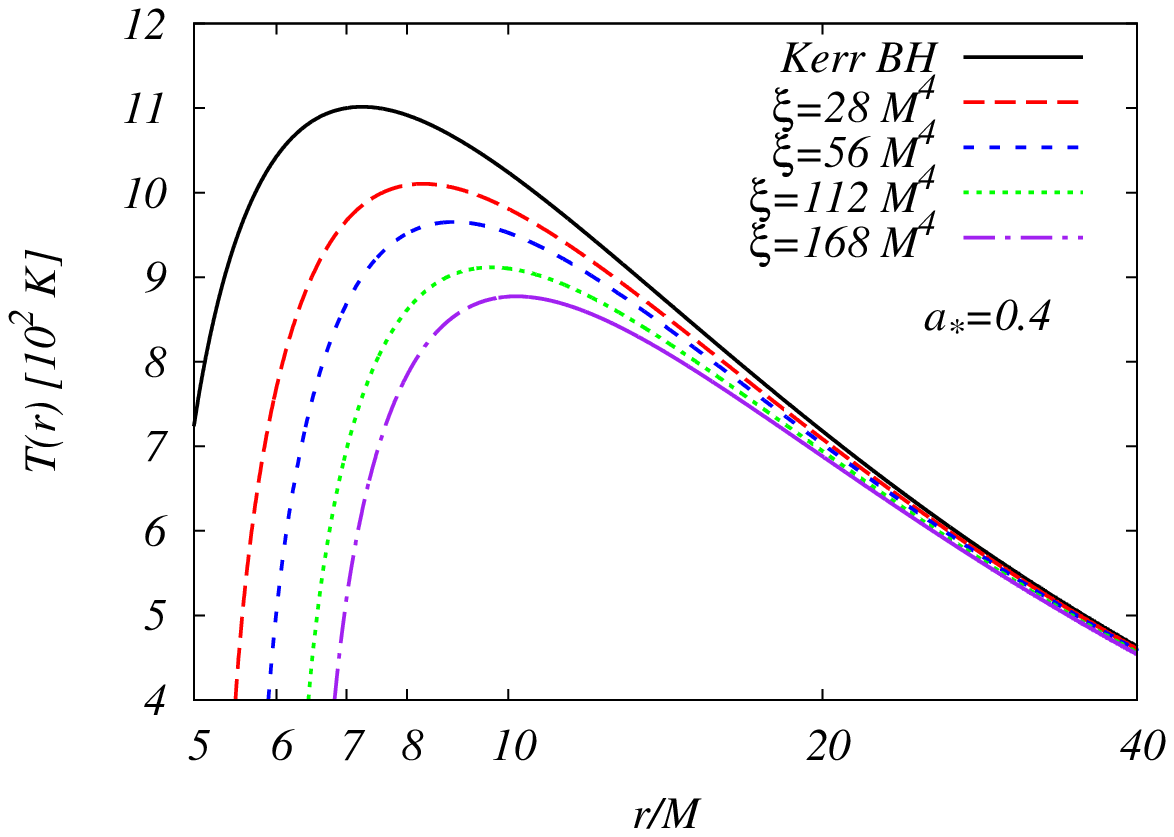}
\caption{The temperature distribution over the accretion disks
around slowly-rotating black holes for different spin parameters
in general relativity and in the modified CS theory of gravity
with the YP approximation. The values of $M$, $a_*$ and
$\dot{M}_0$ are the same as in Fig.~\ref{Fig:flux2}.} \label{Fig:temp2}
\end{figure*}

Similar features can be found in Fig.~\ref{Fig:temp2}, where we
plot the temperature profiles of the disk. Although the
differences here are not so large, since the temperature is
proportional to $F^{1/4}$, the CS approximations of the
slowly-rotating black hole metrics can still be discriminated from
the standard general relativistic case.

\subsection{Disk spectra and conversion efficiency}

\begin{figure*}
\centering
\includegraphics[width=2.8in]{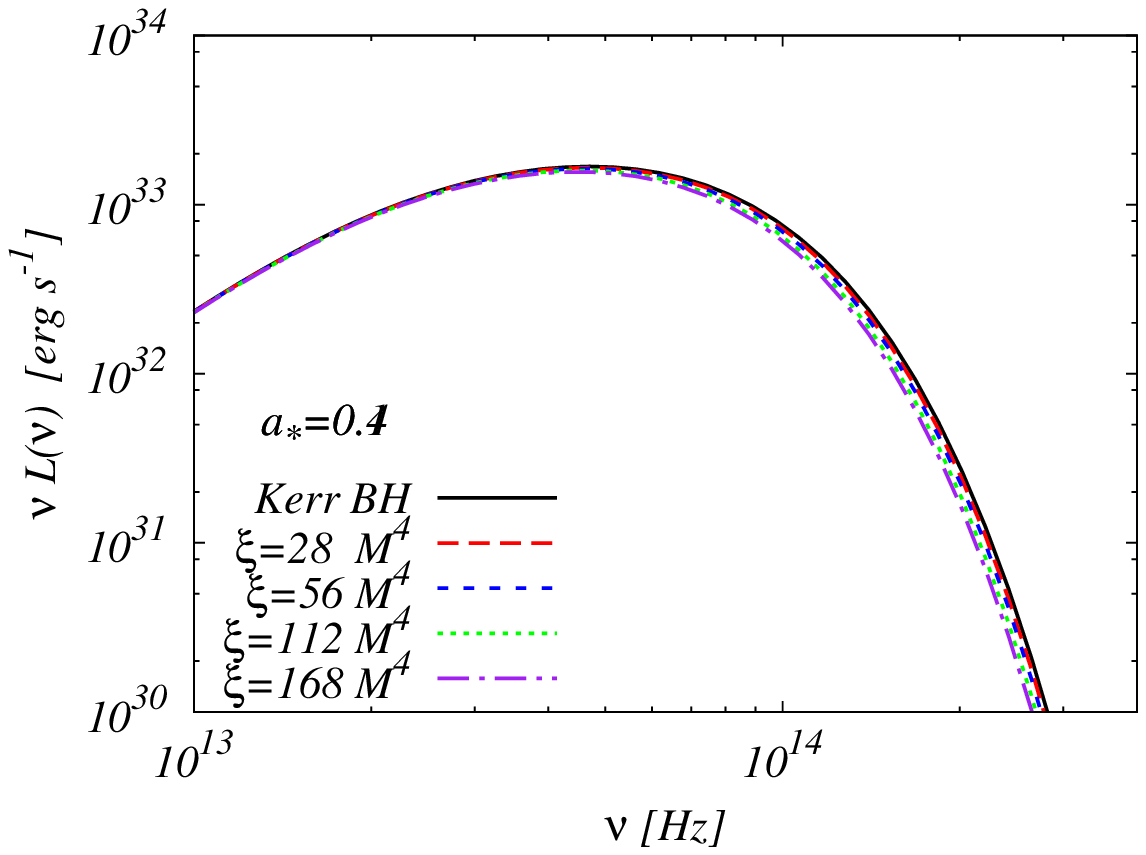}
\hspace{0.2in}
\includegraphics[width=2.8in]{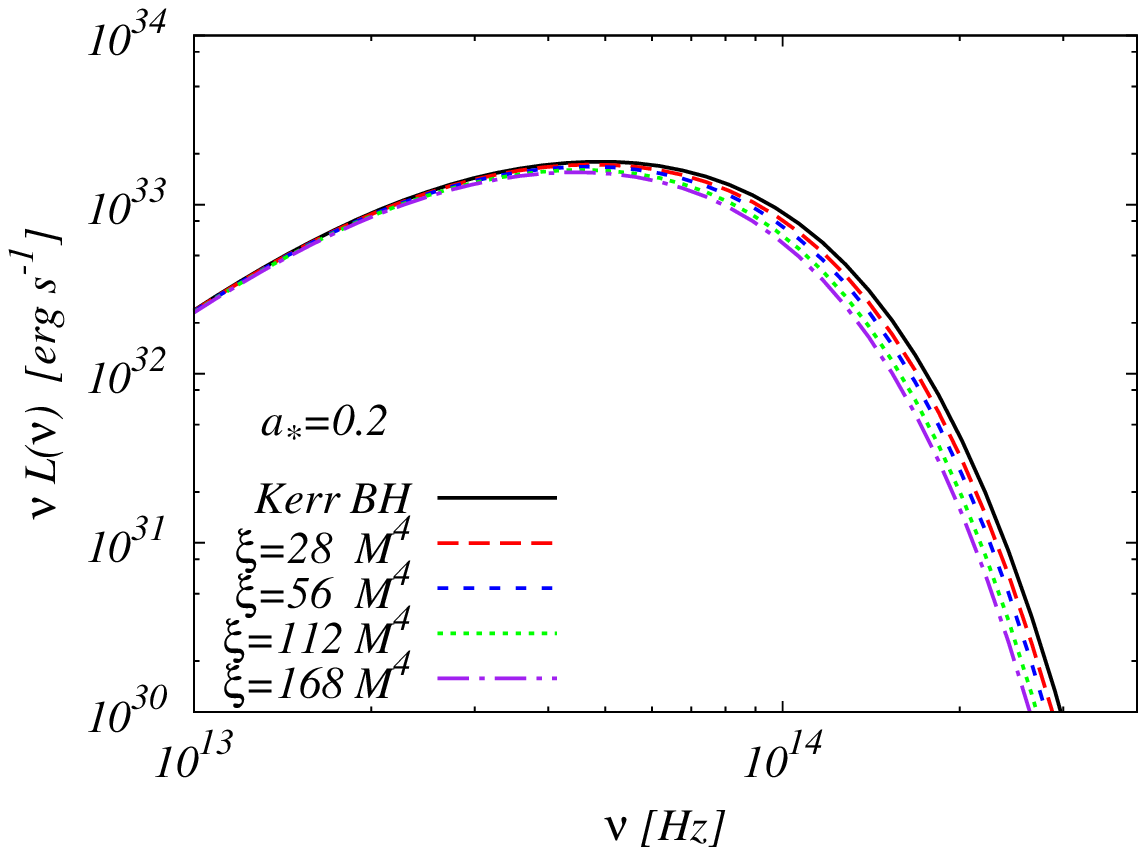}
\hspace{0.2in}
\includegraphics[width=2.8in]{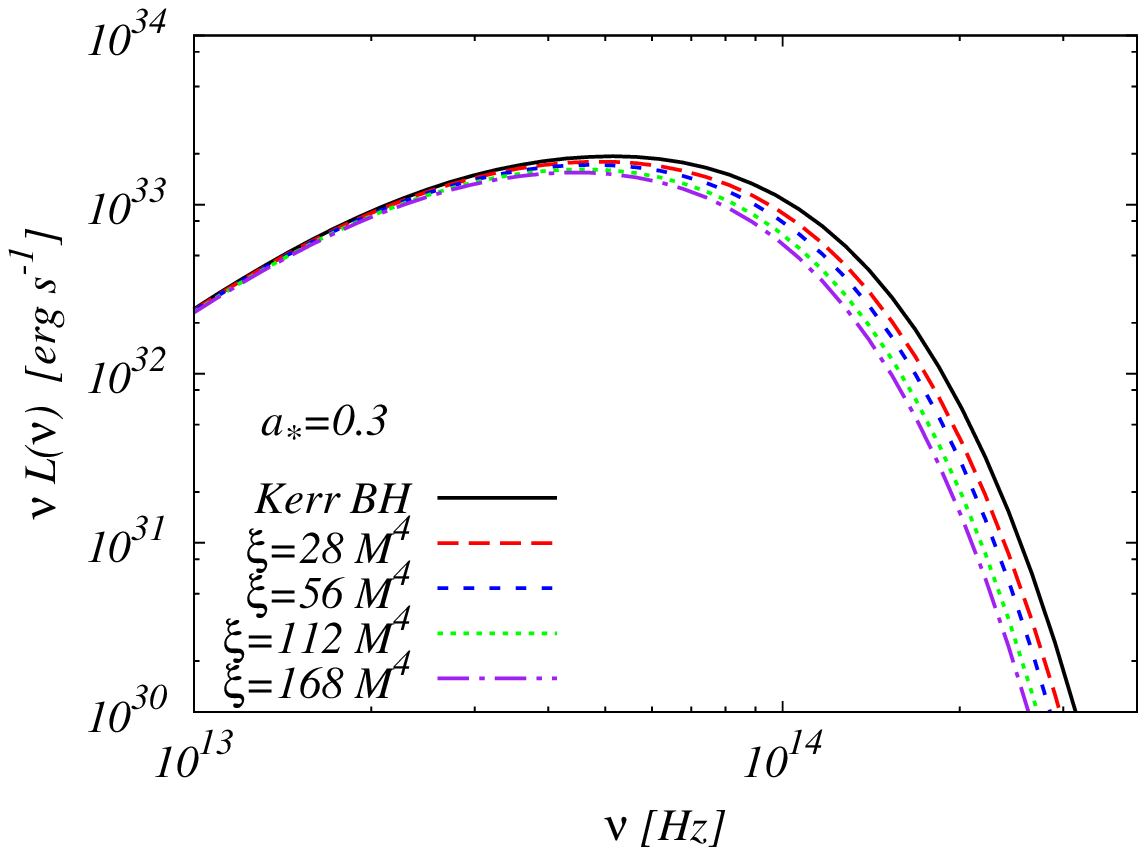}
\hspace{0.2in}
\includegraphics[width=2.8in]{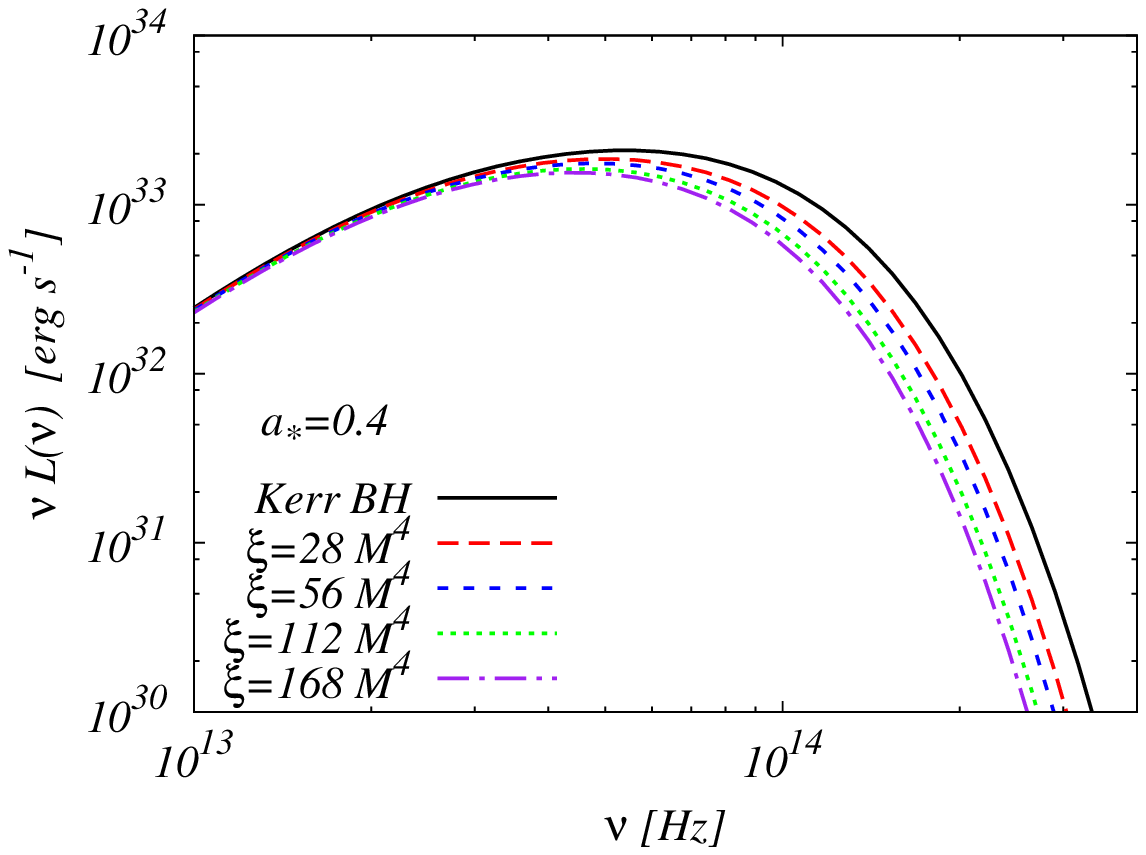}
\caption{The accretion disks spectra for slowly-rotating black
holes with different spin parameters in general relativity and in
the modified CS theory of gravity with the YP approximation. The
values of $M$, $a_*$ and $\dot{M}_0$ are the same as in Fig.~\ref{Fig:flux2}.} \label{Fig:spect2}
\end{figure*}

In Fig.~\ref{Fig:spect2}, we present the spectral energy
distribution of the disk radiation around the slowly-rotating
black holes for the general relativistic case, and for the CS
modified gravity.

The plots show that with the increase of the coupling constants of
the CS gravity, the cut-off frequency of the spectra decreases,
from its value corresponding to the Kerr black hole, to lower
frequencies of the order of $10^{14}$ Hz. Similarly to the case of
the flux profiles, the effects of the CS coupling on the spectral
cut-off are stronger for black holes rotating faster than for very
slowly-rotating black holes. For the radiation of the accretion
disks around black holes this means that the CS theory produces
rather similar disk spectra as in standard general relativity,
even that with increasing coupling constants the radial
distributions of the fluxes differ to some extent for these two
theories (see the top left hand plot in Figs.~\ref{Fig:flux2}, and
\ref{Fig:spect2}, respectively).

\begin{figure*}
\centering
\includegraphics[width=2.8in]{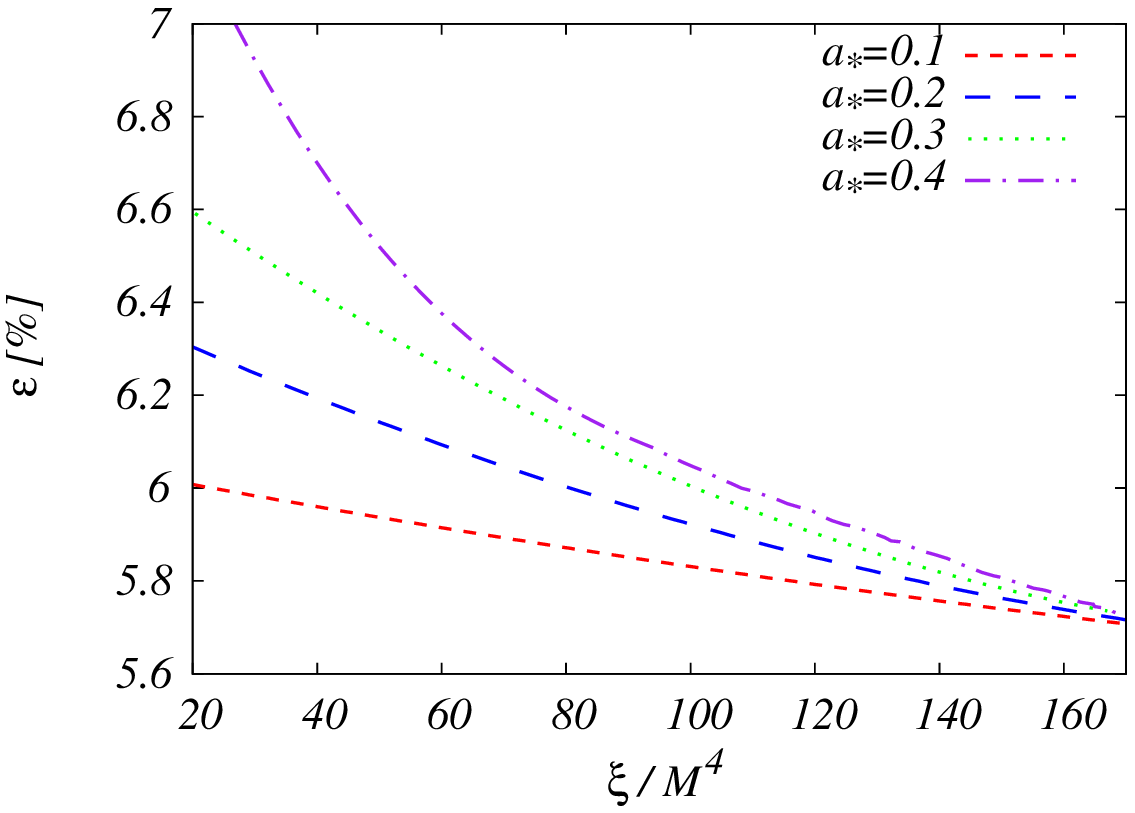}
\caption{The conversion efficiency $\epsilon $ as a function of
the CS coupling parameter $\xi /M^4$ for different values of the
spin parameter.} \label{eps}
\end{figure*}

In Table~\ref{Tab:eff2} we also present the conversion efficiency
$\epsilon$ of the accreting mass into radiation for the case when
the photon capture by the rotating black hole is ignored. The
variation of $\epsilon $ as a function of $\xi /M^4$ is presented
in Fig.~\ref{eps}.

The value of $\epsilon$ measures the efficiency of energy
generating mechanism by mass accretion. The amount of energy
released by matter leaving the accretion disk and falling down the
black hole is the binding energy $\widetilde{E}(r)|_{r=r_{ms}}$ of
the black hole potential.

\begin{table}[tbp]
\begin{center}
\begin{tabular}{|c|c|c|c|}
\hline
$a_*$ & $\xi/M^4$ & $r_{in}/M$ & $\epsilon$\\
\hline
0.1   & -    &  5.6739  &  0.0606\\
      &  28 &  5.7936  &  0.0599\\
      &  56 &  5.8952  &  0.0592\\
      &  112 &  6.0762  &  0.0581\\
      &  168 &  6.2250  &  0.0571\\
\hline
0.2   &  -   &  5.3315  &  0.0646\\
      &  28 &  5.6122   &  0.0626\\
      &  56 &  5.8226  &  0.0611\\
      &  112 &  6.1405  &  0.0588\\
      &  168 &  6.3807  &  0.0572\\
\hline
0.3   &  -   &  4.9818  &  0.0694\\
      &  28 &  5.4662  &  0.0653\\
      &  56 &  5.7744  &  0.0628\\
      &  112 &  6.1951  &  0.0595\\
      &  168 &  6.5025  &  0.0573\\
\hline
0.4   & -    & 4.6182   &  0.0751\\
      &  28 & 5.3546   &  0.0679\\
      &  56 & 5.7456   & 0.0643\\
      &  112 & 6.2550   &  0.0601\\
      &  168 & 6.6100   &  0.0574\\
\hline
\end{tabular}
\end{center}
\caption{The inner edge of the accretion disk and the efficiency
for slowly-rotating black holes in general relativity and in the
CS modified theory of gravity with the YP approximation. The lines
where the value of $\xi$ is not defined correspond to the general
relativistic case. } \label{Tab:eff2}
\end{table}

Table~\ref{Tab:eff2} shows that $\epsilon$ is always higher for
rotating general relativistic black holes than for their
counterparts in CS modified gravity. As the Kerr black holes spin
up, the accreted mass-radiation conversion efficiency raises from
about 6\%, the characteristic value of the mass accretion of the
static black holes, to 7.5\%. This feature is much more moderate
for the rotating black holes in the CS theory: with increasing
rotational velocity, $\epsilon $ also increases; however, the rate
of this increase becomes smaller for stronger CS coupling.
However, these values show that the Kerr black holes provide a
more efficient engine for the transformation of the energy of the
accreting mass into radiation than their slowly-rotating
counterparts in the modified CS theory of gravity, no matter what
approximation is used.

\section{Discussions and final remarks}\label{sec:concl}

In the present paper we have considered the basic physical
properties of matter forming a thin accretion disk in
slowly-rotating black hole spacetimes in the context of the
dynamical formulation of CS modified theories of gravity. The
physical parameters of the disk -- energy flux, temperature
distribution and emission spectrum profiles -- have been
explicitly obtained for several values of the coupling constant
for the YP solution.  Due to the differences in the spacetime
structure, the CS black holes present some very important
differences with respect to the disk properties, as compared to
the standard general relativistic Kerr case. We have also shown
that the Kerr black holes also provide a more efficient engine for
the transformation of the energy of the accreting mass into
radiation than their slowly-rotating counterparts in the modified
CS theory of gravity.

It is generally expected that most of the astrophysical objects
grow substantially in mass via accretion. Recent observations
suggest that around most of the active galactic nuclei (AGN's) or
black hole candidates there exist gas clouds surrounding the
central far object, and an associated accretion disk, on a variety
of scales from a tenth of a parsec to a few hundred parsecs
\cite{UrPa95}. These clouds are assumed to form a geometrically
and optically thick torus (or warped disk), which absorbs most of
the ultraviolet radiation and the soft x-rays. The most powerful
evidence for the existence of super massive black holes comes from
the very long baseline interferometry (VLBI) imaging of molecular
${\rm H_2O}$ masers in the active galaxy NGC 4258 \cite{Miyo95}.
This imaging, produced by Doppler shift measurements assuming
Keplerian motion of the masering source, has allowed a quite
accurate estimation of the central mass, which has been found to
be a $3.6\times 10^7M_{\odot }$ super massive dark object, within
$0.13$ parsecs. Hence, important astrophysical information can be
obtained from the observation of the motion of the gas streams in
the gravitational field of compact objects.

The flux and the emission spectrum of the accretion disks around
compact objects satisfy some simple scaling relations, with
respect to the simple scaling transformation of the accretion rate and mass.
In order to analyze the scaling properties of the physical parameters of the accretion disks we introduce the scale invariant dimensionless coordinate $x=r/M$. Then the functions $h_1(x)$ and $h_3(x)$ given by Eqs.~(\ref{eq:h1YP}) and (\ref{eq:h3YP}) depend only on the scale invariant dimensionless spin parameter $a_*=a/M$, and they have no explicit dependence on the mass $M$. As a consequence, in Eqs.~(\ref{rotE})-(\ref{rotOmega}), only the specific angular momentum and the rotational frequency have an explicit dependence on $M$, in the form $\widetilde {L}\propto M$ and $\Omega\propto M^{-1}$, whereas $\widetilde{E}$ does depend only on $a_*$. For any rescaling $M_2=\alpha M_1$ of the mass of the star we obtain $x_2 = \alpha^{-1} x_1$, and the relations
\begin{equation}
 \widetilde{E}\left(M_2;x_2\right)=\widetilde{E}(M_1)(x_1),
\tilde{L}\left(M_2;x_2\right)=\alpha\tilde{L}\left(M_1;x_1\right),
\end{equation}
and
\begin{equation}
\Omega \left(M_2;x_2\right)=\alpha^{-1}\Omega \left(M_1;x_1\right),
\end{equation}
respectively. For any rescaling $\dot{M}^{(2)}_0=\beta \dot{M}^{(1)}_0$ of the accretion rate, these relations give
\begin{equation}
F\left(M_2,\dot{M}^{(2)}_0\right)\left(x_2\right)= \frac{\beta }{\alpha^2} F\left(M_1,\dot{M}^{(1)}_0\right)\left(x_1\right),
\end{equation}
as the scaling relation of the flux integral given by Eq.~(\ref{F}). Then the temperature scales like
\begin{equation}
T\left(M_2,\dot{M}^{(2)}_0\right)\left(x_2\right)= \left(\frac{\beta }{\alpha^2}\right)^{1/4}
T\left(M_1,\dot{M}^{(1)}_0\right)\left(x_1\right).
\end{equation}
For the maximum of the luminosity $L$ we have $\nu_{max}\propto T$, which gives
$L\left(\nu_{max}\right)\propto \nu^{3}_{max}$. As the frequency scales like the temperature, we obtain that
\begin{equation}
L\left(M_2,\dot{M}^{(2)}_0\right)\left(\nu_2\right)= \left(\frac{\beta }{\alpha^2}\right)^{3/4} L\left(M_1,\dot{M}^{(1)}_0\right)\left(\nu_1\right),
\end{equation}
with
\begin{equation}
\nu_2\left(M_2,\dot{M}^{(2)}_0\right)= \left(\frac{\beta }{\alpha^2}\right)^{1/4} \nu_1\left(M_1,\dot{M}^{(1)}_0\right).
\end{equation}

On the other hand, the flux is proportional to the accretion rate
$\dot{M}_{0}$, and therefore an increase in the accretion rate
leads to a linear increase in the radiation emission flux from the
disk. For a simultaneous scaling of both the accretion rate
$\dot{M}_0$ and of the mass of the black hole $M$, the maximum of
the flux scales as $F\rightarrow\left(\dot{M}_0/M^2\right)F$, with
all the other characteristics of the flux unchanged. Thus, if for
a black hole mass with mass $M_1=10^6 M_{\odot}$ and by
considering a mass accretion rate of $\dot{M}_0^{(1)}=10^{-12}
M_{\odot}$/year=$4.22 \times10^{-10}\dot{M}_{Edd}$, in the case
$a_{*}=0.1$ and $\xi =28M^4$, the maximum of the flux is
$F_{max}^{(1)}=4\times 10^7\;{\rm erg}\;{\rm s}^{-1}\;{\rm
cm}^{-2}$. In the case of a black hole with mass $M_2=10M_{\odot}$
and with an accretion rate $\dot{M}_0^{(2)}=10^{-4}
M_{\odot}$/year=$4.22\times10^{-2}\dot{M}_{Edd}$, the maximum
position of the flux is
$F_{max}^{(2)}=\left[\left(\dot{M}_0^{(2)}/M_2^{(2)}\right)/\left(\dot{M}_0^{(1)}/M_2^{(1)}\right)\right]F_{max}^{(1)}=4\times10^{25}\;{\rm
erg}\;{\rm s}^{-1}\;{\rm cm}^{-2}$. The scaling law of the
temperature is
$T\rightarrow\left(\dot{M}_0^{1/4}/\sqrt{M}\right)T$. Due to the
temperature scaling, the maximum value of the spectrum increases,
but the relative positions of the different spectral curves does
not change.

The determination of the accretion rate for an astrophysical
object can give a strong evidence for the existence of a surface
of the object. A model in which Sgr A*, the $3.7\times 10^6
M_{\odot }$ super massive black hole candidate at the Galactic
center, may be a compact object with a thermally emitting surface
was considered in \cite{BrNa06}.  Given the
very low quiescent luminosity of Sgr A* in the near-infrared, the
existence of a hard surface, even in the limit in which the radius
approaches the horizon, places a severe constraint on the steady
mass accretion rate onto the source, ${\dot M}\le 10^{-12}
M_{\odot}$ yr$^{-1}$. This limit is well below the minimum
accretion rate needed to power the observed submillimeter
luminosity of Sgr A*, ${\dot M}\ge 10^{-10} M_{\odot}$ yr$^{}$.
Thus, from the determination of the accretion rate it follows that
Sgr A* does not have a surface, that is, it must have an event
horizon. Therefore the study of the accretion processes by compact
objects is a powerful indicator of their physical nature. However,
up to now, the observational results have confirmed the
predictions of general relativity mainly in a qualitative way.
With the present observational precision one cannot distinguish
between the different classes of compact/exotic objects that
appear in the theoretical framework of general relativity
\cite{YuNaRe04}.

However, important technological developments may allow one to
image black holes and other compact objects directly
\cite{BrNa06}.  In principle, detailed
measurements of the size and shape of the silhouette could yield
information about the mass and spin of the central object, and
provide invaluable information on the nature of the accretion
flows in low luminosity galactic nuclei.

The spectrum in black hole systems can be dominated by the disc
emission \cite{Do}. Recently, the RXTE satellite has provided a
large number of data of the X-ray observations of the accretion
flows in galactic binary systems. there is also a huge increase of
the radio data for these systems. The behavior of the spectrum in
such systems is consistent with the existence of a last stable
orbit, and such data can be used to estimate the black hole spin.
At high luminosities these systems can also show very different
spectra \cite{Do}. Changes in the spectra of the disks are driven
by a changing geometry. Presently there exists an enormous amount
of data from the X-ray binary systems which can be used to test
this assumption. Therefore the study of accretion processes in Low
Mass X-ray Binaries with well constrained thermal spectra could
also lead to the possibility of discriminating between the various
extensions of standard general relativity.

Hence the study of the accretion processes by compact objects is a
powerful indicator of their physical nature. Since the energy
flux, the temperature distribution of the disk, the spectrum  of
the emitted black body radiation, as well as the conversion
efficiency show, in the case of the Chern-Simons theory vacuum
solutions, significant differences as compared to the general
relativistic case, the determination of these observational
quantities could discriminate, at least in principle, between
standard general relativity and Chern-Simons gravity, and
constrain the parameters of the model.

\section*{Acknowledgments}

We would like to thank the two anonymous referees whose comments
and suggestions helped us to significantly improve the manuscript.
The work of T. H. was supported by the General Research Fund grant
number HKU 701808P of the government of the Hong Kong Special
Administrative Region.


\end{document}